\documentclass[electronic]{vgtc}             %

\usepackage{mathptmx}
\usepackage{graphicx}
\usepackage{times}
\DeclareGraphicsExtensions{.pdf,.png,.jpg}

\ifpdf%
  \pdfoutput=1\relax                   %
  \usepackage{graphicx}                %
  \DeclareGraphicsExtensions{.pdf,.png,.jpg,.jpeg} %
\else%
  \usepackage{graphicx}                %
  \DeclareGraphicsExtensions{.eps}     %
\fi%

\graphicspath{{figures/}{pictures/}{images/}{./}} %

\usepackage{microtype}                 %
\PassOptionsToPackage{warn}{textcomp}  %
\usepackage{textcomp}                  %
\usepackage{mathptmx}                  %
\usepackage{times}                     %
\usepackage{cite}                      %
\usepackage{tabu}                      %
\usepackage{booktabs}                  %

\onlineid{0}

\vgtccategory{Research}

\vgtcinsertpkg

\title{A Graphical Workflow Exploration Environment For Visual Analytics}

\author{Chunggi Lee \thanks{e-mail: $\{$cglee, juyoung0, dryjins, sako$\}$@unist.ac.kr}\\ %
        \scriptsize UNIST %
\and Juyoung Oh $^\ast$\\ %
     \scriptsize UNIST %
\and Seungmin Jin $^\ast$\\ %
    \scriptsize UNIST 
\and Isaac Cho \thanks{e-mail: isaac.cho@usu.edu}\\ %
     \scriptsize Utah State University
\and Sungahn Ko $^\ast$\\ %
     \scriptsize UNIST }

\abstract{Graphical history mechanisms have been widely utilized in many domains to support humans' limited working memory, error recovery, collaboration, and presentation in visual analysis. Yet, there are aspects that remain under-explored in designing graphical history systems for visual analytics systems to help analysts who have complicated workflows. In this paper we report on our design study performed with domain experts, where we characterize domain tasks and designed a visual graphical workflow management environment. Our environment allows analysts to efficiently review, edit, navigate, and explore their complex workflows with their colleagues. In order to evaluate the environment, we present a case study and user study. In the case study, we explore how two domain experts perform collaborative review, communication, and training with our environment; while in the user study with the car data, we reveal that how our environment helps users and how the history mechanism affects users' visual problem-solving behaviors.} %

\teaser{
 \centering
 \includegraphics[width=1.0\textwidth]{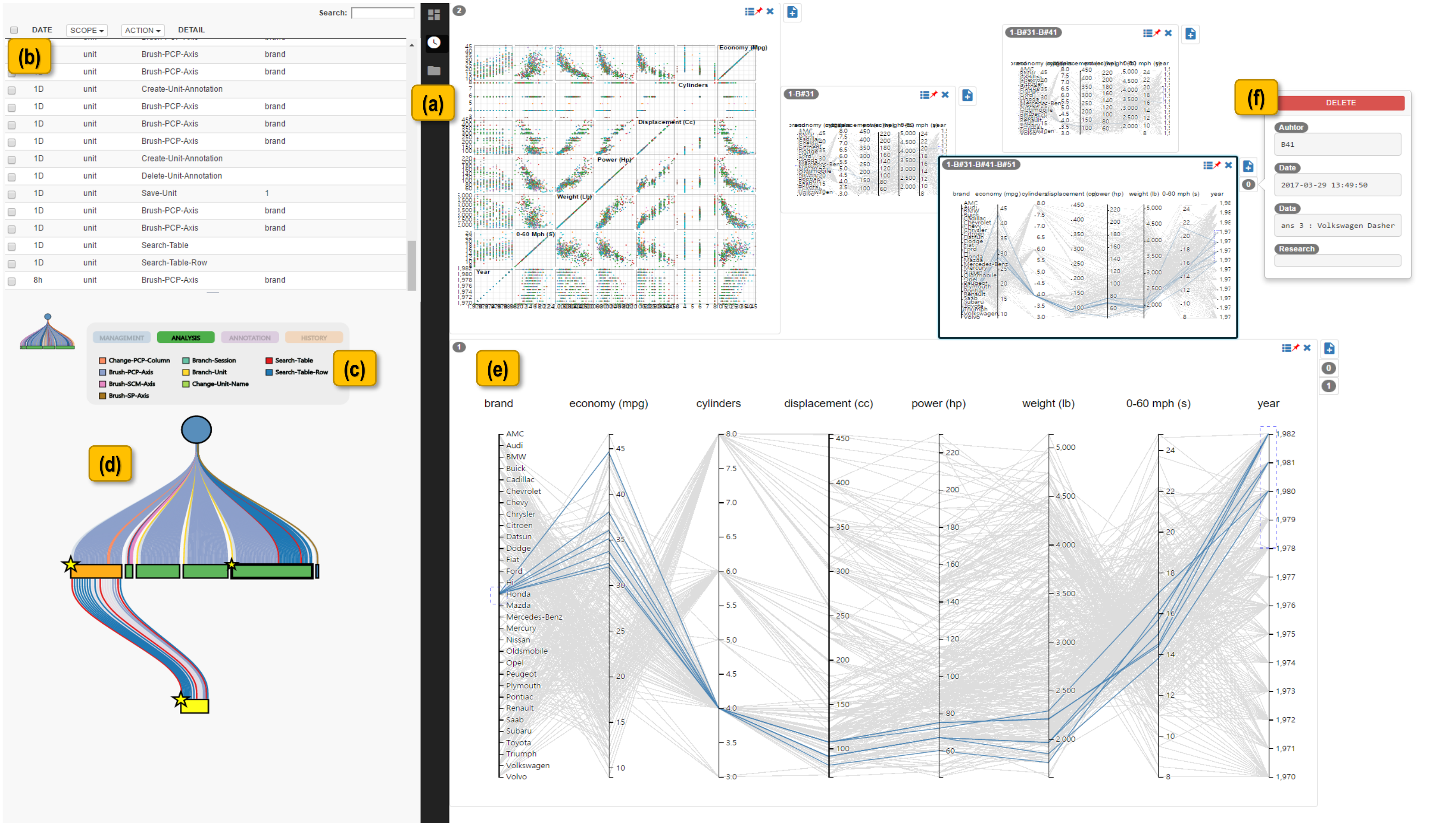}
  \caption{Our graphical workflow exploration system for visual analytics. (a) a menu bar, (b) an action history view, (c) legends for action taxonomy, (d) a unit/session workflow view, (e) a unit with PCP incorporated, and (f) an annotation of a unit. Note that multiple units are presented simultaneously for side-by-side comparison.}
  \label{fig:teaser}
}

\begin{document}

\maketitle

\newcommand{\IconProject}{%
  \begingroup\normalfont
  \includegraphics[height=\fontcharht\font`\B]{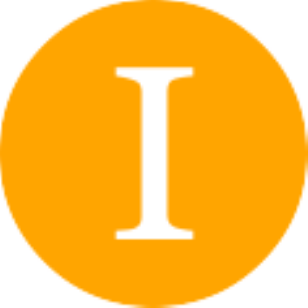}%
  \endgroup
}

\newcommand{\IconSession}{%
  \begingroup\normalfont
  \includegraphics[height=\fontcharht\font`\B]{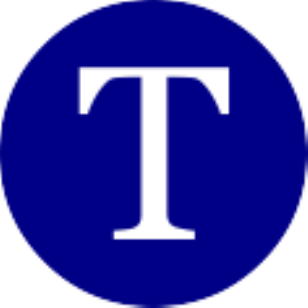}%
  \endgroup
}

\newcommand{\IconUnit}{%
  \begingroup\normalfont
  \includegraphics[height=\fontcharht\font`\B]{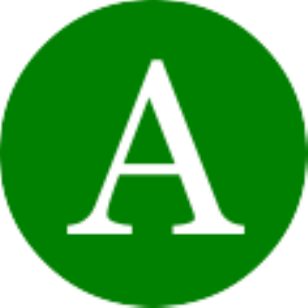}%
  \endgroup
}

\newcommand{\IconWorkflow}{%
  \begingroup\normalfont
  \includegraphics[height=\fontcharht\font`\B]{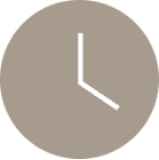}%
  \endgroup
}

\newcommand{\IconGSEA}{%
  \begingroup\normalfont
  \includegraphics[height=\fontcharht\font`\B]{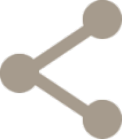}%
  \endgroup
}

\newcommand{\IconGraph}{%
  \begingroup\normalfont
  \includegraphics[height=\fontcharht\font`\B]{./Figures/graph}%
  \endgroup
}

\section{Introduction}
Visual exploration~\cite{Keim02} (or exploratory data analysis~\cite{Tukey77}) is a data-driven approach for deriving and maximizing insights. %
In the exploration, analysts perform multiple rounds of hypothesis generation and validation, each of which leads to the next round with new findings. While many visual analysis tools have improved the quality of analyses, analysts still need more aids due to humans' limited (visual) working memory~\cite{Miller56, Alvarez04}, irregular analysis time spans, error recovery, review, education, collaboration with other analysts~\cite{Saraiya05}, presentation to different levels of management, and reproduction of analyses~\cite{Ragan16}.

Much research has been devoted to solving different aspects of the problem. In order to support error recovery, many models (e.g.,~\cite{Abowd92, Cass06, Berlage94, Derthick01}) have been proposed that are extensively utilized in the image- or photo-editing domains~\cite{Myers15, Chen16, Nancel14}. In addition, the line of research in scientific visualization~\cite{Bavoil05}, visual analysis~\cite{Javed13, Shrinivasan08}, and visual analytics~\cite{Gotz06, Shrinivasan09b, Kadivar09} has become \textit{provenance} research~\cite{Ragan16}, whose systems allow analysts to revisit past visual exploration processes for reviewing, presenting, and recovering reasoning processes~\cite{Lipford10}. In doing so, graphical history interfaces play an important role~\cite{Heer08, Dunne12}. 

However we argue that many aspects remain under-explored in developing a visual analytics system that provides users with history mechanisms. For example, the analysis process in  genome research is very complex~\cite{Nielsen10}, with a large number of comparisons needed for hypothesis testing, reviews, and communications with colleagues~\cite{Saraiya05, Saraiya06}. Yet, few discussions exist on how to (1) manage such complex workflows, which consist of multiple/parallel/hierarchical analysis processes, (2) decide the information granularity for review and communication purposes, (3) resolve dependency issues caused by the modification of analysis pipelines, (4) remove repeative arduous operations~\cite{Saraiya06}, and (5) provide efficient navigation and exploration of the workflows~\cite{Heer08}. Although visual analytics systems that are equipped with history mechanisms have already been requested~\cite{Saraiya06}, rarely have visual tools or guidelines been proposed.

In this paper, we report a design study we performed with two analysts to develop ation environment. During collaboration with the analysts in comparative analysis, we held regular bi-weekly meetings, multiple on-site interviews, and several seminars, in order to formulate task characterization, and design rationales. As a result of the collaboration, we propose a graphical workflow management and exploration environment that can help analysts by providing: (1) efficient navigation and exploration of workflows with a multi-level sankey visualization, (2) both action and state level histories for review and communication, (3) an automated cascading action dependency algorithm, (4) various high level management actions (e.g., copy, move, cut, and paste a series of actions) as well as conventional modification operations (e.g., delete, skip, undo, redo, selective undo~\cite{Archer84, Yang88, Abowd92, Vitter84, Berlage94, Myers15}), and 5) insight provenance processes based on context-aware annotations and workflows. Lastly, we evaluate our environment with two domain experts and a user study with the car data set in order to demonstrate the generality of our environment for aiding users' limited visual working memory. 

To our knowledge, little research has been perfromed for insight provenance~\cite{Pohl10, Gotz08}, where users solve problems with their visual workflows being presented during their visual analysis. Groth and Streefkerk~\cite{Groth06} present a user study that measures the effectiveness of history mechanisms, but this work is limited in that there is only one task, which involves finding the centrality of 3D molecule and is not related to data analytical tasks. In this work we perform a user study with the conventional car data~\cite{Grinstein02}, where users can use both visualizations and their visual workflows. From the workflows of 48 participants, we report an observation that different problem-solving strategies (i.e., branch-based, or history-based) can produce different performance in the tasks with visualizations. The major contributions of this paper are: 

\begin{itemize}
  \item We present systematic task abstraction and challenges in the domain as well as user tasks in order to derive the design goals and rationales for our environment, 
  \item We design an efficient graphical workflow environment with an handler for finding broken pipelines,
  \item We evaluate the proposed environment with a case study, a user study, and domain expert feedback, and
  \item We demonstrate with workflows left by users that different user preference affects on visual problem-solving pattern and performance.
\end{itemize}

\section{Related Work}
There are multiple components for designing a graphical workflow management, including history models, data structures, undo/redo functions, action editing with dependency checks, and error recovery. The research on history mechanism becomes popular with a series of algorithmic discussions for error recovery~\cite{Archer84}, redo~\cite{Vitter84}, undo support~\cite{Yang88, Abowd92}, and selective undo~\cite{Berlage94}. However, these techniques do not consider visual interfaces of recent systems. Thus, they have been utilized for text-based tools, such as text editors.   

As interactive visual interfaces become proliferated, the research focus moves to developing new history algorithms and models for the visual interfaces~\cite{Myers96} or collaborative work environments~\cite{Edwards00}. In the visualization community, Kreuseler et al.'s work~\cite{Kreuseler04} can be considered one of the early work that tries to utilize the history mechanism for visual data mining. In the work, many fundamental concepts are considered that are also utilized in this work, including operations, dependency, management of history. Scientific visualization systems also exploit the history mechanism. Ma demonstrate that rendered images can be better managed as a form of a graph. Followed by Ma's work, Bavoil et al.~\cite{Bavoil05} also propose an image-based multiple-view visualization system, where the pipeline concept for rendering is described. Later, Heer et al. provide a comprehensive survey on the history mechanisms and demonstrate an graphical interface incorporated in Tableau. 

Our work is built on top of all the previous work and discusses more considerations, in order to design an efficient graphical workflow exploration environment for visual analytics systems. To our knowledge, little research has been performed on how to develop history mechanisms for visual analytics systems. In this work, we provide design rationales to utilize history mechanisms for helping limited visual working memory. In doing so, we also present a new approach for visually exploring and editing workflows and handling broken pipelines caused by user editing.

\subsection{Insight Provenance with Interaction History}
Visual analytics is the science of analytical reasoning facilitated by interactive visual interfaces \cite{cook2005illuminating}. One key aspect that separates visual analytics from other related fields is the focus on analytical reasoning. While the final products generated from an analytical process are of great value, research has shown that the processes of the analysis themselves are just as important if not more so \cite{pike2009science}. Such processes not only contain information on individual findings, but also how users arrive at the findings. This area of research that focuses on understanding a user’s reasoning process through the study of their interactions with a visualization is referred to as Analytic Provenance. Dou et al. have defined 5 stages in order to summarize and understand analytic provenance. The 5 stages include: perceive, capture, encode, recover, and reuse. Perceive is related to how data is being presented to the user; while capture relates to capturing the provenance by interaction logs, annotations, video recording, eye-tracking, etc. Prior work has captured and analyzed user interactions \cite{dou2009recovering}, and found that strategies related information can be extracted from interaction logs. Encode refers to the process of describing the captured provenance in predefined formats for further analysis. Researchers have attempted using XML \cite{jankun2007model}, declarative pattern language \cite{xiao2006enhancing}, and dynamic scripts \cite{kadivar2009capturing}, but in most cases these schemas only record the ``how'', but not always the ``why''.  Recover is the next stage of extracting insights from the encoded provenance. Finally, reuse is related to being able to reapply a user's insight to a new dataset. Kadivar et al. \cite{kadivar2009capturing} have presented a visual analytic system that turns the user interaction logs into a script that can be reused on different datasets.

\section{Domain Characterization}
\label{sec:DomainCharacterization}
Bio-data are dynamic networks of interacting components, and the complex interactions between the genome, proteome, and metabolome define biological functions. Recent bio-data have some common characteristics, such as high-throughput and multiomics. 
We collaborated for 10 months with two analysts, one from a corporation and another from a genome research center. The main business model of the corporate is producing predictive personal health-care reports based on recent genome research, while that of the research center is publishing scientific journals. During the collaboration, we held bi-weekly meetings and occasionally half- or full-day seminars for iterative system design~\cite{Sedlmair12}. We limited our topics to gene expression analysis because the scale of genome research is too broad to cover~\cite{Nielsen10}; the analysis include pathways, browsers, assemblies, sequences, and alignments. In this section, we first characterize and abstract~\cite{Munzner09} task requirements in comparative genomics, a well-known comparative analysis field~\cite{Nielsen10, Seo02}, and the main challenges derived from the domain experts' experience. 

\begin{figure}[t]
    \begin{center}
	\includegraphics[width=1.0\columnwidth]{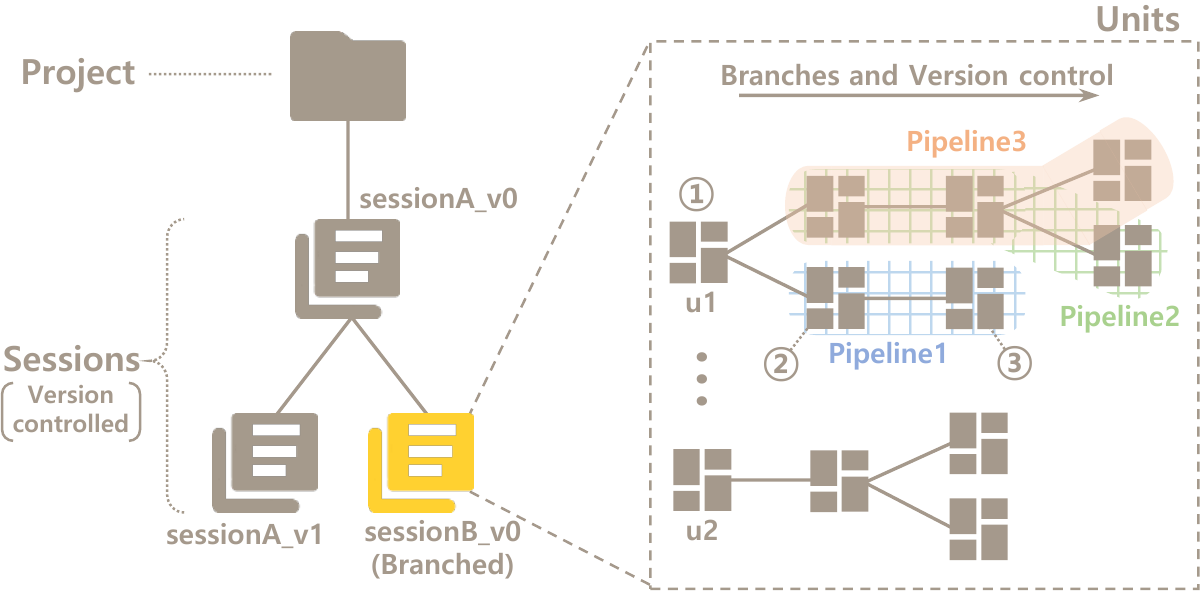}
	\caption{An example workflow in visual analytics. A Workflow can be consisted of multiple version-controlled sessions that include various pipelines in different versions for analysis.}
	\label{fig:workflow}
	\end{center}
	\vspace{-0.7cm}
\end{figure}

\subsection{Problem Formulation and Task Requirements}
\label{sec:ProblemFormulation}
Genome research can be characterized by its very large and complex data for analysis in nature and thus requires different workflows~\cite{Nielsen10}. Therefore, many tools have been proposed for analysts in the area. %
In order to determine the advantages and disadvantages of current tools, we reviewed 28 visual tools for genome alignment, 28 visual tools for pathway analysis, 30 genome browsers, 35 comparative genomics tools, 22 gene expression viewers, and other general network visualization tools, such as 21 tree viewers, 50 network visualizations, etc.%

In the first few meetings, the analysts presented their workflows and visual tools. According to the analysts, \textit{a workflow} (Fig.~\ref{fig:workflow}) comprise a number of \textit{projects} (\includegraphics[width=0.03\linewidth]{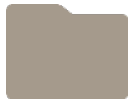}) where  they perform \textit{sessions} (\includegraphics[width=0.03\linewidth]{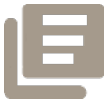}) with groups of data (e.g., experimental data produced in different biological treatments and at different times). Here, a session can be defined as ``an interval of more or less continuous interactive activity"~\cite{Archer84}, and the duration of each session can span hours, days, or a few weeks according to the number and complexity of the visualizations (\includegraphics[width=0.028\linewidth]{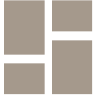}) %

An interesting point to consider is the existence of \textit{analysis pipelines}, each of which consists of a series of operations to produce a meaningful visualization state, also described in VisTrails~\cite{Bavoil05} for analyzing diffusion tensor data. For example, one of their pipelines can (1) load genes to run clustering algorithms, (2) validate genes if they are clustered properly in a biological context, and (3) verify the quality of clustering results with interaction networks. If the quality is not satisfied, the pipeline returns to (1) and chooses different algorithms or parameters. In the discussion, the analysts mentioned that this workflow, which is also described in \cite{Lyi15} is quite general in the field.

After the meetings, we found that analysts' difficulty came not from a lack of visualizations, but from poor support for managing their complex workflows, because many changes, modifications, and extensions of pre-existing pipelines and workflows can happen during self- and peer-review work. For example, among multiple sessions that an analyst performs as shown in Fig.~\ref{fig:workflow}, Pipeline 1 can be described as (1) loading data, (2) selecting a clustering algorithm, and (3) selecting parameters. A few months later, when Pipeline 1 is finished, the analyst needs to test other color schemes in preparing a publication or running many other clustering algorithms due to recently found discoveries in newly published work. This work requires recovering previous workflows, modifying current pipelines, and even developing new pipelines. But doing the work generates entangled actions and pipelines, and is cumbersome as Bavoil et al. point out~\cite{Bavoil05}. Currently, the analysts do not have any support from their visual tools in this respect.

When analysts are assigned to a project with in-house experiment data, they iteratively compare their data to the data in pre-existing databases, such as KEGG~\cite{Kanehisa00}, in multiple sessions, in order to find new knowledge, validate hypotheses, and extract new insights by using visual approaches (e.g., heatmaps and clustering algorithms as shown in Seo's work~\cite{Seo02}). %
During or after a series of analyses, the analyst needs to retain the information from the process they performed (i.e., data, algorithms, parameters), any new insights and knowledge they obtained, and a future analysis direction, as well as management information (i.e., a project name, analysts' name, dates, times, and annotations). However, to the best of our knowledge, there is no system-wide support provided by current visual tools for their recording purposes. Using research notebooks, similar to the diaries described in ~\cite{Saraiya06} is the current method analysts use in their visual exploration.  

Having conclusions is not the end of an analyst's work. Often, the analyst needs to discuss the conclusions with his/her colleagues. In general, each discussion produces a number of suggestions for improving pipelines, which include not only changing algorithms, parameters, and color schemes, but also editing the actions performed in the pipeline or even initializing new sessions with new data for further investigation. Once multiple rounds of discussions have produced more convincing conclusions than earlier conclusions, the director joins the discussion in order to review and confirm them for use in scientific publications. Here, many recommendations for improvements are again provided, this time from the director with which the analyst should revisit and edit the analysis pipelines. The problem here is that it can be very hard for the analyst to recall details of the pipelines that she performed because discussions might occur a few days, weeks, or even months after analyses. The only strategy that the analysts can use is to write down every analytical step in as much detail as possible, but such writing requires a large amount of effort and time. Our analysts reported that they do not have any visualization tools to support not just their workflows, but also annotations critical for both individual and collaborative analysis~\cite{Mahyar10, Mahyar12} %

In summary, due to a lack of appropriate support from current tools for comparative analyses, analysts have difficulties at many stages (e.g., during and after their analysis, in communicating with colleagues for improving analysis pipelines for better results, and for recovering errors). Here, we summarize and formulate the problems as \textit{task requirements} for a comparative analysis environment as follows:

\begin{itemize}%

\item[R1]\textbf{Complete history for both actions and states:} Users should be able to revisit any state or action for recalling and reviewing previous analyses. 

\item[R2]\textbf{Hierarchical workflow history management:} All information in a workflow should be presented, navigated, and managed hierarchically for multiple levels of analysis for efficient work management and communication. 

\item[R3]\textbf{Flexible, editable, and reusable workflow:} A workflow model and its data structures should be flexible enough to support hypothesis testings and allow for editing actions for recovering errors and improving the quality of analysis pipelines. Users should be notified if their edit breaks the original pipelines.
 
\item[R4]\textbf{Visual workflow exploration environment:} Workflows are complicated and can span over long periods of time. An environment should provide a sufficient visual context for the workflows and allow for their easy navigation and exploration.

\item[R5]\textbf{Comparative analysis support:} The primary data dimension of interest in this work is comparative genome analysis; for this reason, the environment should support efficient methods for this work.%

\end{itemize}

\begin{figure*}[t]
  \centering
  \includegraphics[width=1.0\linewidth]{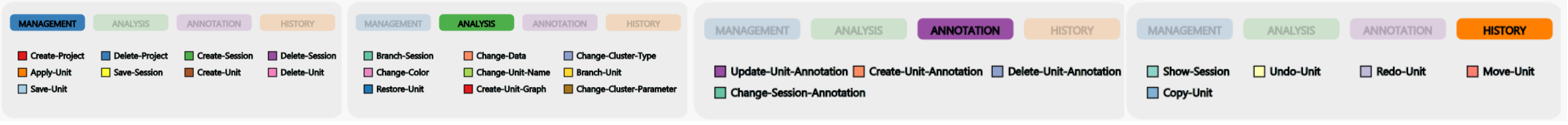}
  \caption{Our action taxonomy. There are four action categories: management, analysis, annotation, and history. This taxonomy helps analysts see what actions are performed on each unit and session.}
  \vspace{-0.3cm}
  \label{fig:actionTaxonomy}
\end{figure*}

\section{Design Rationales for a Visual Workflow Exploration Environment}
\label{designRationales}
With the formulated domain characterization and requirements in Section~\ref{sec:DomainCharacterization}, we performed a survey for analyzing algorithms, models, and approaches related to history mechanisms. Then, we extracted design considerations for a visual workflow exploration environment that can be used for visual analytics systems. In this section, we present our design rationales with background information, in terms of visualization units, data structures, models, and visual workflow. 

\subsection{Unit-based Comparative Analysis}
\label{sub:unit}
Utilizing multiple views allows side-by-side comparisons, which in turn allow an efficient analysis. However, if the number of views is not determined with caution, a usability problem and inefficient space allocation are easily generated. In the comparative analysis, the number of views during an analysis is often unclear and hard to decide upon, depending on the number of data sets, goals and analysis pipelines. This means there should be a flexible means for visual comparative analysis (R5). 

Our environment provides \textit{units} (e.g., Fig.~\ref{fig:teaser} (e) and  Fig.~\ref{fig:workflow} \includegraphics[width=0.032\linewidth]{./Figures/unit.pdf}) %
that can be independently visualized, filtered, exported, bookmarked, and modified as suggested in~\cite{Kang12} for data investigation. Using the units implies that all the information and actions relevant to a unit is independently saved to a data structure attached to the unit, including the data sets used and actions performed for changing unit's properties (e.g., size, position, color scheme), as well as other analytical options (e.g., algorithms, parameters, annotations). However, such recordings can be tricky without further consideration of hierarchically-structured workflows, because units could grow non-linearly for comparisons or version-control. For example, a unit can be branched for separating workflows (e.g., ``sessionA\_v1'' and ``sessionB\_v0'' for hypothesis testing in Fig.~\ref{fig:workflow}) or saved with progress (e.g., ``sessionA\_v0'' and ``sessionA\_v1'' units in Fig.~\ref{fig:workflow}). Here, branching a unit means creating a new unit with the same history of the unit being branched. Branches tend to imply insight generation or hypothesis testing during analysis and are, therefore, good candidates for future reviews. Lastly, annotations can be made for each unit that play an important role in collaborative reviews~\cite{Mahyar10}. Note that when there is an annotation, a star is assigned as shown in Fig.~\ref{fig:workflow}. In the next section, we present our considerations, in order to support history mechanisms for unit-based hierarchical workflows. 

\subsection{Models and Data Structures for Workflow History}
There is much research on what should be captured for building a history between \textbf{actions} and \textbf{states}. In our work, a state is defined as \textit{``the settings of interface widgets and the application content''} and an action is defined as \textit{a series of events that ``transform one state into another''} ~\cite{Heer08}. The action-based model provides a highly detailed history of an analysis and tends to be combined with undo/redo operations for editing~\cite{Chen16, Myers15}. But an information overload and scalability problem can be generated without appropriate presentation and navigation methods for actions while using this model. On the other hand, the state-based model captures the final state with surrounding variables and parameters generated by a series of actions during visual exploration (e.g., \cite{Dunne12, Ma99}). The state-based model assumes the use of visualizations for presenting states and has been popular with multiple visualization views. But this model alone is not enough, because pipeline reviews and improvements require action-level review and modification as well (R3).

\begin{figure}[t]
    \begin{center}
	\includegraphics[width=0.9\columnwidth]{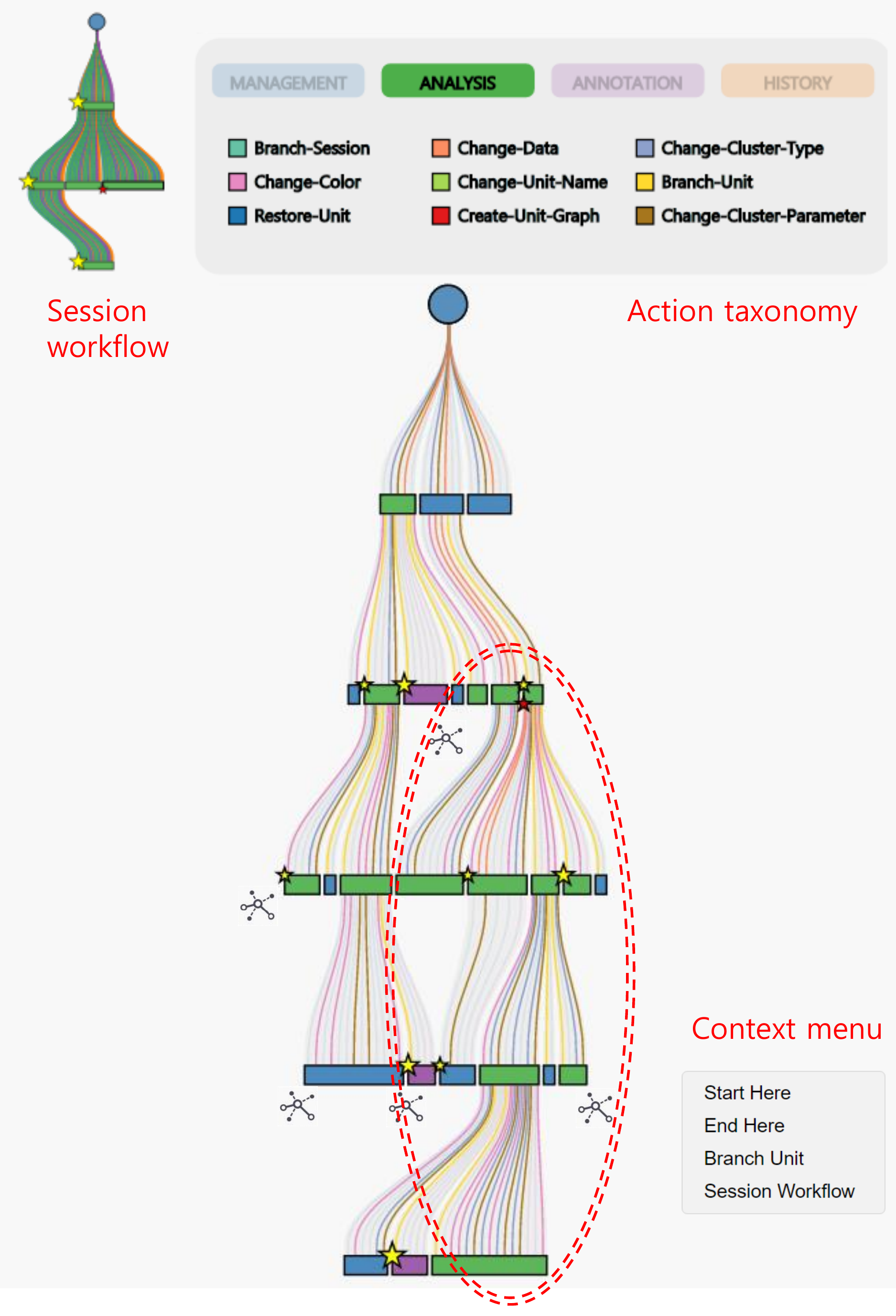}
	\caption{An example of a session/unit workflow view. (top-left) An overview of a session workflow, (middle) sankey visualization for a unit workflow. The stars represent annotations.  Different colors of lines between units represent newly performed actions. The color of each unit is a representative color that describes the actions on the unit. (top-right) A legend for our action taxonomy. Note that most nodes are green. This implies that the most frequently performed actions in this session are analysis actions.}
	\label{fig:unitSankey}
	\end{center}
	\vspace{-0.8cm}
\end{figure}

The next consideration is data structure. At first, we found that using a linear list is common and easily allows for the maintenance of an action history with undo or redo operations~\cite{Abowd92}. This structure, however, is ineffective for review purposes, because recent actions in the history are lost after backtracking (i.e., unable to maintain a complete history (R1)~\cite{Cass06}). As an alternative, a tree structure for maintaining a complete history~\cite{Berlage94, Derthick01} has been utilized in many history systems, because it naturally accommodates hypothesis testing with branches, where both original and controlled data can be compared simultaneously. 

In our environment, analysts can have both state-level visual overviews (i.e., time-machine computing~\cite{Rekimoto99}), and action-level pipeline reviews (e.g., selection of algorithms, parameters, undo, redo, insert, cut, move). In doing so, we utilize both the linear list and tree structure and provide both ``local and global history''~\cite{Heer08}. With the local history, analysts are able to review and modify individual actions that are saved in each unit's linear list, while the histories of all the units can be reviewed and managed by the session-level global history with the tree structure. In the next section, we describe how to visually explore the workflows with branches and restore analysis states, branches, and actions. 

\subsection{Visual Workflow Management and Exploration}
\label{sub:Sankey}
Unorganized comparative analysis workflows cause problems in the management of analyses. For example, when an analyst needs to review work performed a long time ago, she might have to rely on partial and incomplete information from the work, including data, dates, and analysis types, each of which is mundane and hard to recall. Conventional history systems use the thumbnail-based presentation approach~\cite{Terry04, Grossman10, Chen11b, Ma99} for helping users recall, but the approach is not easily adapted without considering the modification and navigation of both actions and states (R3 and R4). According to the analysts, little support is provided from their visual tools for workflows management or for the exploration of their workflows. Therefore, research notebooks are still extensively used, which require much more effort and time to maintain. 

In order to allow visual hierarchical workflow management (R2), our environment utilizes a sankey diagram \cite{riehmann2005interactive} for presenting the workflows in sessions and units, as shown in Fig.~\ref{fig:unitSankey}. Note that currently the top-left overview is the session workflow, while the middle sankey visualization presents the workflows of the units in one of the sessions. 
In the diagram, parent-child or sibling relationships exist between two neighboring nodes. The links between a parent node and a child node reveal the actions performed on the child node after it is created. The color of each unit node is from the taxonomy (Fig.~\ref{fig:actionTaxonomy}), in order to represent the dominant action on each unit.

The workflow diagram provides rich interactions in order for efficient workflow exploration. At first the session and unit workflow visualizations are switchable. In the session workflow, users can click any node (i.e., a session) for time-machine computing~\cite{Rekimoto99} that reproduces a previous visual state of units. A list of actions is also presented in the action history view when each node is clicked. For a series of actions performed over multiple nodes, the user can specify a start and end session (orange and yellow colors in Fig.~\ref{fig:teaser}). Branching a new session is also allowed in the diagram to separate workflows. Lastly, analysts can review a detailed session history by selecting a session in the diagram as shown in Fig~\ref{fig:unitSankey}, where multiple units exist.  

The unit workflow diagram shows pipelines of multiple units in a session. This diagram works in the same way that the session diagram works with one difference: the interactive mini overview of the session workflow as shown in Fig.~\ref{fig:unitSankey} (top-left), and a legend for action taxonomy (top-right). We categorize the actions that can be performed in our environment as management (e.g., create a unit), analysis (e.g., change an algorithm), annotation (e.g., creating an annotation), and history (e.g., undoing/redoing) as shown in Fig.~\ref{fig:actionTaxonomy}. With the context menu shown in Fig.~\ref{fig:unitSankey}, users are allowed to switch between a session workflow and a unit workflow and select two units or sessions to list the actions between the two selected nodes.

\subsection{Editable and Reusable Workflows with Dependency}
\label{sub:editWorkflow}
Analysts are regularly asked to review and edit previous workflows for recovering errors and improving pipelines, when better approaches can be found in recent publications. In order to support the tasks and make the analysts comfortable in editing and trying new actions~\cite{Kuttal11}, our environment utilizes a selective undo/redo mechanism~\cite{Berlage94} with an interactive interfaces as shown in Fig~\ref{fig:teaser} (b) and (d). Here, when a user selects a unit for editing in the unit workflow view (d), a list of the actions in the unit is shown in (b) for editing (e.g., undo and redo). Other high level workflow management actions are also provided in our environment, including copy, move, and skip actions, in order to help the analysts reuse previous pipelines~\cite{Kurlander92, Myers96} without performing repeated actions that Saraiya et al.  point out as a problem in their longitudinal study~\cite{Saraiya06}. In our environment, when a unit is dragged and dropped into another unit in (d), all (i.e., copy) or some (i.e., move) of the actions of the selected unit can be copied to the target unit. But providing editable workflows cannot be achieved unless unintended side-effects (e.g., producing semantically ambiguous or meaningless pipelines) are prevented by detecting and handling conflicts as Kreuseler et al. point out~\cite{Kreuseler04}; for example, if deleting the first data loading action in a pipeline causes a broken pipeline. 

\begin{figure}[t]
    \begin{center}
	\includegraphics[width=1.0\columnwidth]{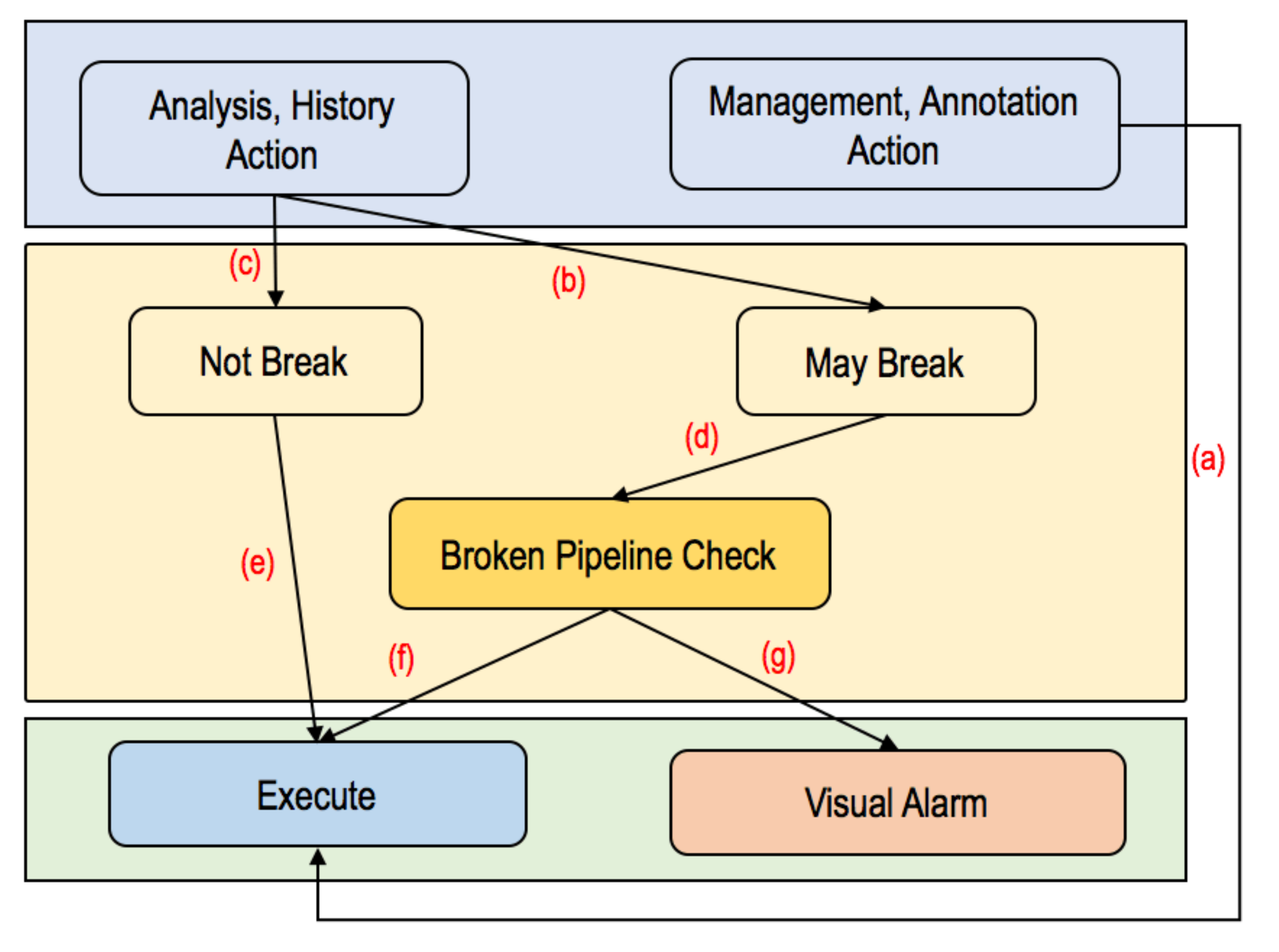}
	\caption{Our approach for finding a broken pipeline caused by a user action. The algorithm first checks the action's category. If the action can cause a broken pipeline, our algorithm searches for an alternative action to suggest. When a pipeline is broken, a visual warning is shown for user awareness.}
	\label{fig:Handler}
	\end{center}
	\vspace{-1.0cm}
\end{figure}

\begin{figure*}[t]
  \centering
  \includegraphics[width=1.0\linewidth]{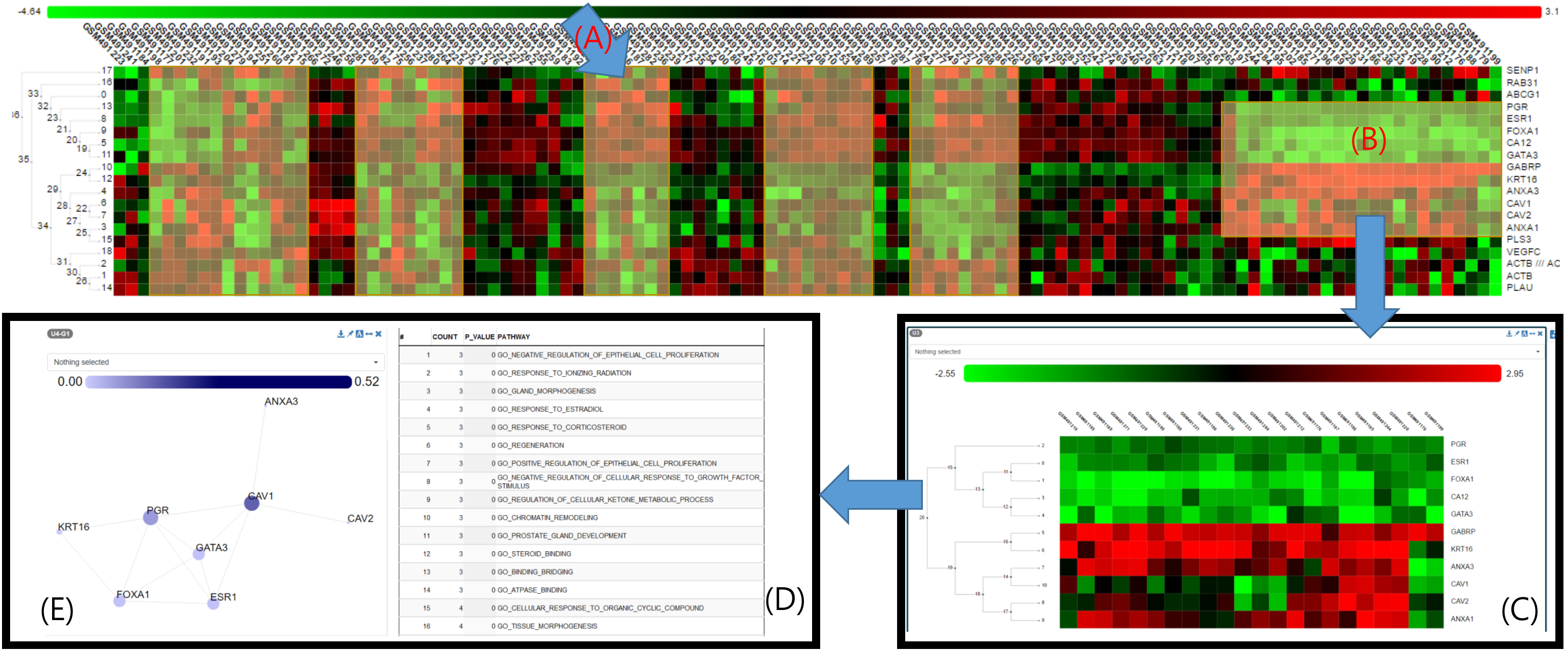}
  \caption{An example of collaborative analysis, communication, and training with our environment. The junior analyst creates five sub-heatmaps with the yellow areas (e.g., (A)) for hypothesis generation. But the senior analyst suggests area (B) that was not spotted by the junior analyst due to less experience. The hypothesis is investigated by GSEA (D) and protein-protein network (E).}
  \vspace{-0.4cm}
  \label{fig:case1}
\end{figure*}

Little attention has been given to the semantics of managing conflicts caused by user edits, since early research proposed simple models for a linear structure~\cite{Berlage94} and scripting applications~\cite{Myers98}. We find that Kreuseler et al.~\cite{Kreuseler04} consider dependency and independency in terms of a chain of (system- or user-level) operators, but little description or semantics is provided. Here, we propose our approach for detecting and handling dependency or conflicts in order to avoid pipeline breaks and provide users with visual awareness of the dependency status. 

When we review previous work, we find that the conventional term of ``dependency" detection or handling is hard to be reused in our environment for two reasons. First, an editing action on a unit has nothing to do with the actions on other units. Second, the final action overrides all the preceding, same types of actions. This means that  only the final action affects the final visual state. Thus, the dependency issue that Cass and Fernades describe~\cite{Cass06} does not happen in our environment. 

But we do find that there are situations where actions can break analysis pipelines. An example case is that after algorithm and parameter selection actions are performed, the user can undo the parameter selection action or copy a series of units as a pipeline to another unit, but it is possible that not all the pipeline units are copied properly. In both cases, a pipeline cannot be run to generate a visualization, becoming a broken pipeline. 

Initially we consider running our detector, whenever an action is performed. But we soon determine that some actions, as well as the final action on the leaf units, have nothing to do with breaking pipelines. For example, creation/deletion of a unit/annotation cannot break a pipeline ((a) in Fig.~\ref{fig:Handler}). In contrast, we find that most of the actions in the analysis category and for editing need the detection process (case (b)). Example actions in this case include the actions for algorithms/parameter changes and undo/redo.

If a detection process is run, all the actions are traversed both upward and downward from on which the unit an action is performed. In a situation where a pipeline can be broken due to an action, the detector searches to find an alternative action to the input action to prevent broken pipelines. If there is an alternative action, our detector provides a visual warning to users that the last action breaks a pipeline and details which previous actions can be taken to prevent the broken pipeline. Here, the user also has an opportunity to undo the last action (f). Otherwise, the detector flags the unit as broken (g).

\section{Considerations for Gene Expression Analysis }
In order to demonstrate the effectiveness of our environment with real-world comparative analysis workflows (R5), we incorporate a heatmap, network visualization, and gene set enrichment analysis (GSEA) to present data. Heatmaps have been extensively utilized in comparative genome analysis~\cite{Saraiya04} due to their scalability--they can visualize more than one thousand genes for an analysis. In addition, the display is easily equipped with automated algorithms, such as clustering algorithms and statistical techniques~\cite{Seo02, Lex10}. In our work, we utilize the heatmap visualization (Fig.~\ref{fig:case1}, top and (C)), combined with several clustering algorithms, including hierarchical, spectral, and k-means algorithms.

As rich information on the interactions between genes, gene products, enzyme reactions, and compounds has been revealed, much research has been performed toward navigating and exploring interconnected pathways and regulatory networks~\cite{Streit08, Westenberg08b, Lambert11, Lex13}. In addition, most visual tools are equipped with network visualizations, supporting knowledge validation between analysts' knowledge and the information in public databases (e.g., KEGG ~\cite{Kanehisa00}). Our environment provides a protein-protein interaction (PPI) network visualization (Fig.~\ref{fig:case1}, bottom-left, (E)) with different sizes for a computed degree of nodes and different colors for the centrality of a node.

Gene set enrichment analysis (GSEA) as shown in Fig.~\ref{fig:case1}, (D)) %
identifies genes that are over-represented, and characterizes biological pathways related to differentially expressed genes. In this study, we analyze the data in the context of several databases such as KEGG~\cite{Kanehisa00} and Gene Ontology (GO)~\cite{Ashburner00}, as suggested by MsigDB v5.2~\cite{Subramanian05}. In order to examine the mapped pathways, we use Fisher's exact test.

\section{Case Study: Collaborative Gene Expression Analysis, Review and Training}
In this section, we evaluate our environment by performing a case study  with two domain experts, a senior (Ph.D., Biology) analyst and junior (M.S., bioinformatics) analyst. In the study, we proceed with the junior analyst first, and then ask the senior analyst to join later, in order to observe how an individual analyst uses our environment with visual workflows and the senior analyst collaborates, communicates, and trains the junior analyst. %

For the study, they use transcriptome profiling data that represent a large amount of quantitative gene expressions produced by using microarray. Microarrays are powerful tools for analyzing gene expression and helping to understand the intricate biological systems involved with normal and diseased organisms. In this work, we use the microarray data that are constructed for determining the expression levels of human breast cancer subtypes. 

In the tutorial session, the analysts are presented with our environment first, and then start using the environment for testing its provided functions, such as sankey workflow visualizations for sessions/units, branches for hypothesis testing, state recovery, version-control, annotations, and the editing actions. 
After the tutorial session, which lasted for about 30 minutes, they are asked to answer the questions taken from Table 7 in Saraiya et al.'s work~\cite{Saraiya05} in order to determine whether the visualizations in our environment are appropriate for their tasks. They answer that they can find answers for questions 1 to 7 from the time series data set with our environment, including changes in overall expression with time, different patterns of expression, and functional details of genes showing high change. When the study session begins, the junior analyst loads 19,615 Human breast cancer data into the environment and creates a heatmap with a hierarchical clustering algorithm. While looking at the heatmap, the analyst often makes a number of heatmaps and PPI networks as well as a gene list in the GSEA view by selecting and branching regions as shown in Fig.~\ref{fig:case1}. During the analysis, we observe an analysis loop as described in ~\cite{Sacha14}. The loop starts with the heatmap visualization, where some regions on the heatmap (e.g., Fig.~\ref{fig:case1}, A) are selected for generating other heatmaps and network visualizations, followed by an  investigation of p-values in the GSEA view, the degree of links, and the centrality of gene nodes. 

At this point, the senior analyst joins the analysis, mentioning that Fig.~\ref{fig:case1} B) is a better candidate region and worth selecting and creating heatmaps, due to the larger fold-change value (i.e., experimental value difference). Initially, the region (Fig. ~\ref{fig:case1} (B) was not spotted by the junior analyst. In order to demonstrate how to efficiently choose areas for hypothesis generation on heatmaps, the senior analyst creates a branch from junior's session workflow, copies a part of the pipelines in the workflow, undo the first data load action, and loads stomach cancer data that she brings. The workflow for this is shown in Fig.~\ref{fig:unitSankey}.

Finishing the training on how to find good candidate areas for hypothesis generation, the senior analyst comes back to the junior's workflow by using the state recovery function. Then she drags the candidate area of (B), which is considered to be related to differentially expressed genes in breast cancer subtypes, and perform clustering again to generate (C) in Fig.~\ref{fig:case1}. As a result, the senior analyst concludes that the green and red block genes in (B) are strongly related to cell proliferation and morphogenesis, both of which play a major rule in the cancer process. Constructing and verifying the PPI network (E) and GSEA (D), the senior analyst finds that ESR1 and PGR genes are the key molecules for identifying the breast cancer subtypes. Other researchers also have confirmed that ESR1, PGR, FOXA, and GATA3 are closely linked to the human breast cancer subtypes, and can help identify the luminal subtype of breast cancer (e.g., ~\cite{Perou00}). At the end of the study, the junior analyst makes annotations about the discussions and lessons, and bookmarks for important sessions and units. Also some heatmaps and networks are exported as vector images for future use.

To sum up, during the session, the senior analyst efficiently collaborates with the junior analyst by utilizing our environment's support. For the purpose of training~\cite{Saraiya06}, which was not intended at the beginning of the process, the senior analyst demonstrates how to find critical regions on heatmaps for yielding better results, which implies improvements of the pipelines. In doing so, both analysts leave their own workflows, as shown in Fig.~\ref{fig:unitSankey} that can be reused for future (collaborative) reviews. Note that the senior analyst's branches and saves for the training are shown in the red ellipse in Fig~\ref{fig:unitSankey}. In addition, we can that the most frequently performed actions are for analysis (green). A pictogram (\includegraphics[width=0.03\linewidth]{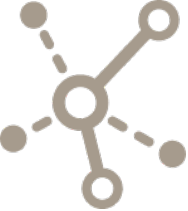}) is used to represent a branch for visualizing the PPI network.

\subsection{Domain Experts' Feedback}
In designing our environment, we have collaborated with the analysts, 
one of whom has extensive experience in the comparative genomics field. They note that although many tools exist for the genomic analysis~\cite{Nielsen10}, little support has been provided for managing and reviewing their pipelines and workflows. In this sense, the support for selecting, branching, and visualizing a subset of the data from heatmap visualizations are very useful.

The junior analyst mentioned that every one or two weeks, new experimental data arrive for analysis. While the data themselves are complicated, the actual difficulty comes from building the same or similar pipelines every time, as described in \cite{Saraiya06}. For example, when there are some areas in the data for detailed comparisons, the analyst has to use R for data processing (i.e., cutting the part with selected dimensions and conditional values) and performs the same actions to build the previous pipelines. In this context, the junior analyst said, the high level editing ability of our environment effectively reduces the burden in the analysis. In addition, reproducing, comparing or improving previous analysis pipelines according to new research results have been difficult tasks for  entry-level analysts. Time-machine computing for state recovering, editing, and recovery abilities in the environment not only make the tasks easier, but also more comfortable ~\cite{Kuttal11}. 

On the other hand, the senior analyst note another aspect. In general, having discussions and collaborative reviews is not an easy task due to diversity in the pipelines, which are difficult to instantly modify for hypothesis generation and testing. One competitive benefit of our environment is the unit-based visualization, which allows analysts to find the list of genes with  side-by-side comparisons with interactive filtering. The analyst said that the task of finding the list can be done with other tools, but that was much easier and quicker with our environment. Overall, the support for easy branching, action editing, and state recovering not only allows time-efficient analysis, but also enables efficient communications and training. The analyst also point out a weakness in terms of less diversity in visualizations with the units. For example, the units can visualize time-series experimental data and present relationships between the data among units, in order to initiate another type of analysis. Lastly, an automated approach for updating the server database is required, due to the high speed of new knowledge generation in the field.

\begin{figure*}[t]
    \begin{center}
	\includegraphics[width=.85\linewidth]{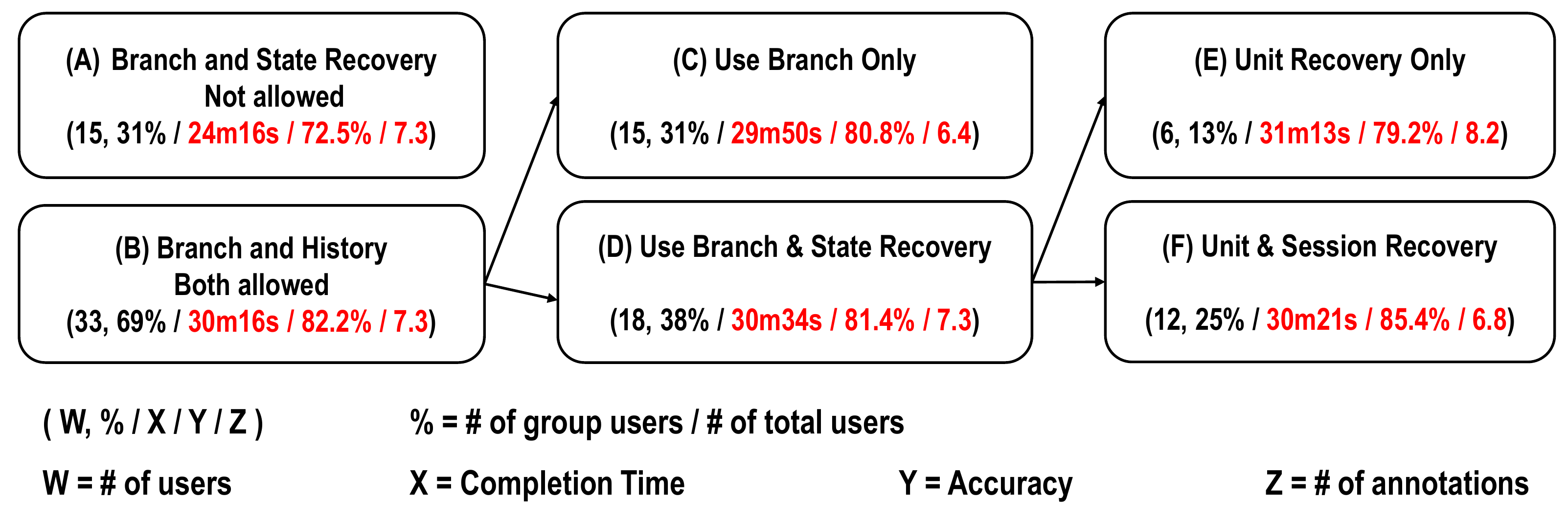}
	\caption{Our study design and result. We initially assign the participants into two groups, one without any functional support (group A) and another with both state recovery and branch functions (group B). During analyzing the result, we find that group B can be divided into two sub groups based on the participants' preference on the history mechanism in problem-solving. Note that even with the participants who prefer to using the history mechanism can be separated based on their interactions between those who use unit recovery only and the others using session and unit recovery.}
	\label{fig:studyResult}
	\vspace{-0.4cm}
	\end{center}
\end{figure*}

\section{User Study}
In this section, we report a user study that demonstrate our graphical workflow environment with units can be used for visual analyses and insight provenance. To have generality, we utilize the car data set~\cite{Grinstein02}. In terms of tasks and questions, we initially complied from previous work~\cite{Amar05a, Valiati06, Kobsa01}. But we soon find in our pilot study that the conventional questions are not appropriate, because many of them can be easily solved without using branches and history mechanisms, the two benefits of our environment. Thus, we decide to create new questions as shown in Table~\ref{fig:questions} that can be efficiently solved with our environment. Note that they are related to each other. With Q3--Q5, we expect to users to recover previous states for aiding the limited working memory in their visual analysis.

\begin{table}[t]
    \caption{We add three analytical questions (Q3~Q5), in addition to conventional compound task questions (Q1~Q2), in order to find the effectiveness of our environment with units, branches, and state recovery.}
	\begin{center}
	\includegraphics[width=0.9\columnwidth]{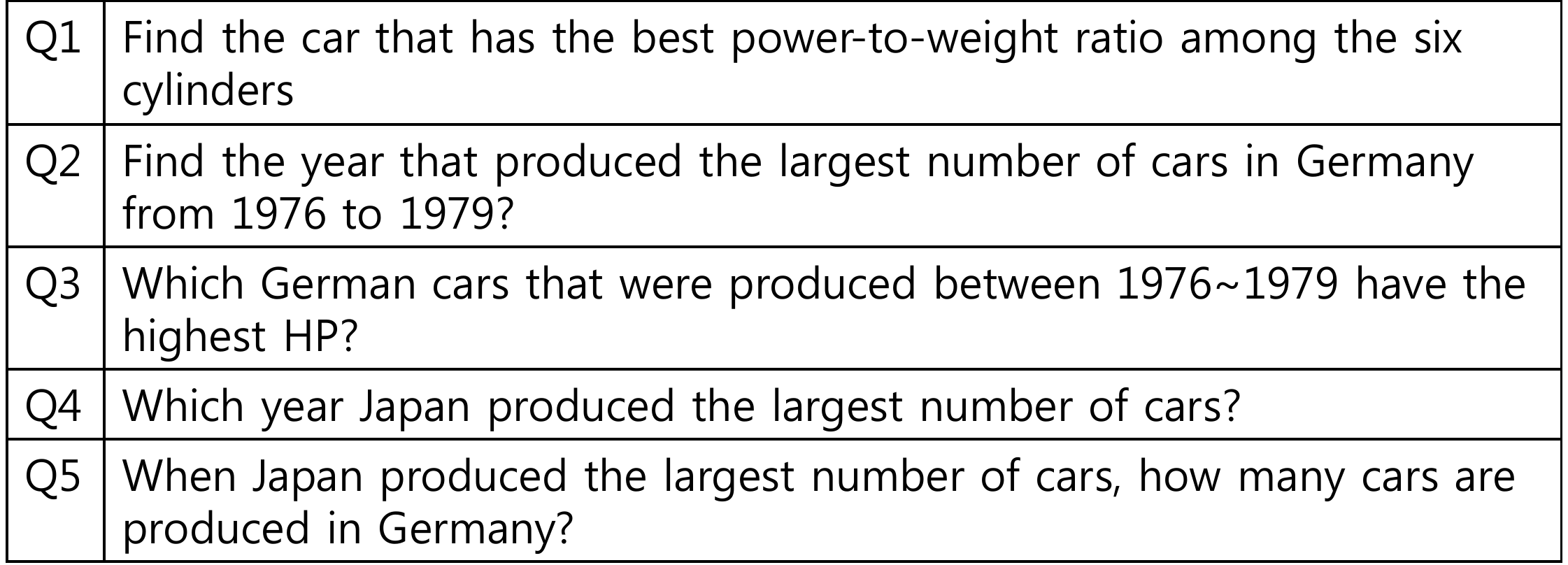}
	\label{fig:questions}
	\vspace{-0.8cm}
	\end{center}
\end{table}

\begin{figure*}[h]
  \centering
  \includegraphics[width=1.0\linewidth]{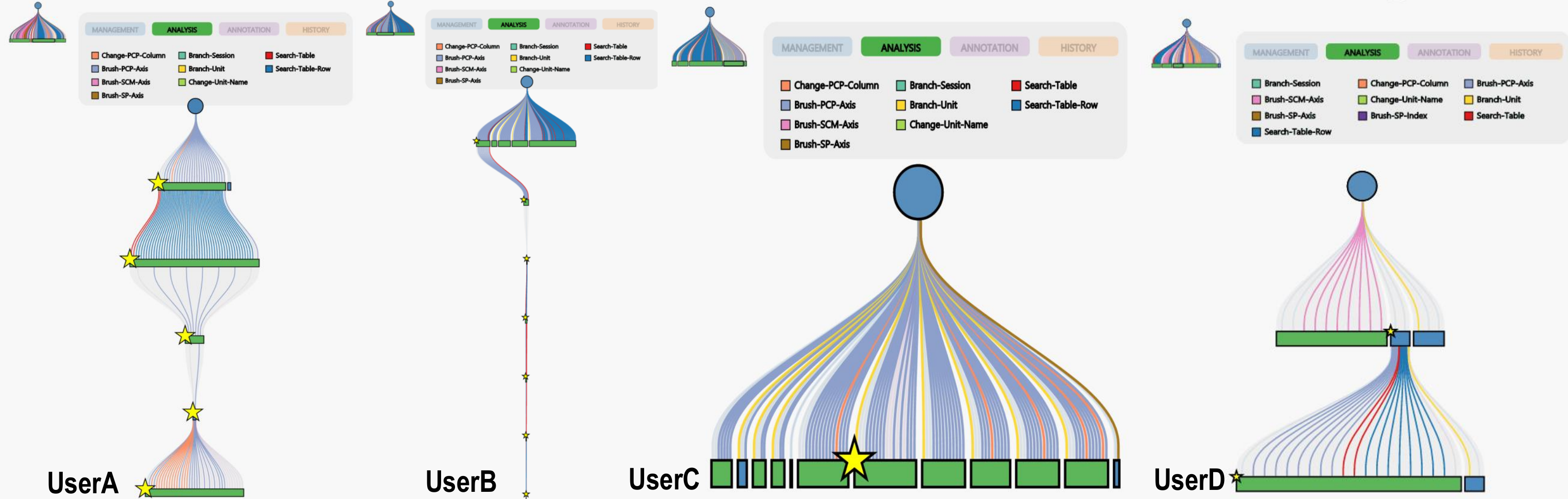}
  \caption{Different workflows according to users' preference in visual problem-solving. Here, the workflows of UserA and UserB imply that the users frequently save units and use the saved unit with the state recovery function for problem-solving (Group E). UserC's workflow means that the user prefers to making branches for side-by-side comparisons (Group C). UserD's workflow represents that the users use both the branch function and state recovery (Group F). Note that userA answers questions in each session, while userC writes all the answers in one unit (i.e., a large star). }
  \vspace{-0.6cm}
  \label{fig:workflowPatten}
\end{figure*}

In terms of visualizations, we incorporate a parallel coordinate plot (PCP, Fig.~\ref{fig:teaser} (e)) and scatterplot matrix (SPM, Fig.~\ref{fig:teaser} (a)) visualizations into the units in our problem-solving environment ~\cite{Grinstein02}. Compared to conventional PCP and SPM techniques, there is a difference in the visualizations. With our environment, both PCP and SPM can be branched for side-by-side comparisons with the interactions performed on the origin unit being remained on the new unit. For example, the user is allowed to move any axes in PCP in a unit (unitA). If a branch unit from unitA is created after an axis is moved to another location, the newly created unit (unitB) also shows the axis in the moved location in unitA. 

Participants were recruited via advertisement emails.
The participants receive a \$6 gift card for the participation. Before the experiment, the participants are asked their demographics (e.g., gender, age, education level), and their familiarity of visualization. Next, three questions are given to participants for exercises, They can ask questions to an instructor while solving the exercises. An example problem-solving process with PCP and SPM is shown in Fig.~\ref{fig:teaser}. After the exercises, the participants are asked to solve five tasks. Then, all the participants are requested to fill out a post questionnaire that is examined for finding the preferences of the system on a 7-point Likert scale. Overall, the duration of the experiment does not exceed 1 hour. 

In order to measure the effectiveness of branching and state recovery in general visual analysis, we create two groups (Group A and Group B) of users. The participants in Group A (15 participants) do not have any functional support, while those in Group B (33 participants) are allowed to use both branching and session/unit state recovery. At this point, the participants of Group B are instructed to create and save a session for questions, so that they can reuse the saved sessions for state recovery, if necessary. Next we present our study result. 

\subsection{Result}

We recruit 48 participants (9 females and 39 males, 43 undergraduate and 5 graduate students). Their average age is 23.3 years old, ranging from 19 to 27. The participants are not really familiar with visualizations (3.8 out of 7). 

Fig.~\ref{fig:studyResult} shows the result of our experiment in terms of the speed and accuracy. After the experiment, we find that Group B can be divided into two sub groups (Group C and Group D) according to problem-solving preferences. Group C mostly uses the branch function, while Group D utilizes both the branch and state recovery ability. We can  further divide Group D into two groups, Group E, whose participants prefer using the unit state recovery capability and Group F, whose participants use both session and unit recoveries. In the next session we present our analysis. We use a 2-sample t-test to check if our experiment result obtained is significant ($p<0.05$) within a 95\% confidence interval. 

We find that Group A solves the questions faster than Group B ($t(46)=1.91, p<0.003$), but has a lower accuracy that Group B ($t(46)=1.93, p<0.029$). We conjecture that providing a systematic support of branches and history mechanism helps users to perform more comparisons. Such concentration of problem-solving produces the higher accuracy. 

An interesting point is that when we analyze the output of the experiment, we find that participants show different preferences in visual problem-solving. For example, because we do not instruct the participant to use a specific approach between using branches and visual history mechanism, the participants decide the approach they use. During our data analysis, as summarized in Fig.~\ref{fig:studyResult}, we see that 31\% of the participants (Group C) in group B mainly use the branch function (i.e., they do not make many saves for state recovery in the future), while the others (Group D) rely on both branches and state recovery mechanism. In addition, as we investigate the interactions of Group D, we find that even Group D can be further divided with two groups, a group who mainly uses unit state recovery and another group who utilizes both unit and session recovery in their problem-solving. Although using branches (Group C) seems to require less time with less accuracy, compared to group F, we do not find any statistical significance between the two groups.

Investigating users' interaction history, we see that different individual preferences in problem-solving leave different visual workflows. For example, in Fig.~\ref{fig:workflowPatten}, we first recognize that the workflows of UserA and UserB grow downward, due to their frequent saves. As revealed in Fig~\ref{fig:studyResult}, this type of behavior helps users utilize the state recovery function for problem-solving (Group D, Fig.~\ref{fig:workflowPatten}). Compared to this, we can verify that UserC and UserD mostly use branches in their problem-solving. With these visual workflows left, we conjecture that visual analytics systems need to consider their users' preferences in problem-solving, if possible, because the preference may cause unexpected performance problems.

\section{Conclusion and Future Work}
In this work, we present a design study for providing analysts with a graphical workflow exploration environment, whose visual workflows help not only visual analysis, but also collaborative reviews and improvements of previous analyses. Our environment provides a means to hierarchically manage the workflows and records user actions during analysis that are presented back to the analysts for their review, edits, and exploration. In addition, our environment is flexible in that the units can be incorporated with other visualizations and tasks. Lastly, we perform a case study with domain experts that presents how our environment helps domain experts' collaborative reviews, communications, discussions, and training. In the user study, we use the conventional car data for demonstrating generality of our environment. In addition, in the experiment, we find that using branches or history mechanism both help users visual analysis. 

As future work, we think that formulating interaction models among units helps enhancing comparability of the units~\cite{North00} for other domains' exploration with different visualizations. In addition, it is an interesting question that how much visual workflows can help users recall and solve problems. For this, we plan to perform an extended user study with the participants who attended the user study in this work, in order to analyze if the left visual workflows can help users' recall and problem-solving.

\bibliographystyle{abbrv}
\bibliography{template}
\end{document}